\documentclass[usenatbib]{mnras}
\usepackage{natbibmnfix, graphicx, times, amsmath, epsfig, amssymb}
\usepackage{verbatim}
\usepackage{color}
\usepackage{mathptmx}
\usepackage{txfonts}
\usepackage[T1]{fontenc}
\usepackage{aecompl}

\def\simgt{\lower.5ex\hbox{\gtsima}} 
\def\simlt{\lower.5ex\hbox{\ltsima}} 
\def\gtsima{$\; \buildrel > \over \sim \;$} 
\def\ltsima{$\; \buildrel < \over \sim \;$}
\def\Msun{M_\odot}

\newcommand\lsim{\mathrel{\rlap{\lower4pt\hbox{\hskip1pt$\sim$}}
        \raise1pt\hbox{$<$}}}
\newcommand\gsim{\mathrel{\rlap{\lower4pt\hbox{\hskip1pt$\sim$}}
        \raise1pt\hbox{$>$}}}
\def\myputfigure#1#2#3#4#5%
{\vskip#5pt\makebox[0pt]{\hskip#2in
\includegraphics[width=#3\textwidth]{#1}}\vskip#4pt\hfill}

\title[First Identification of DCBH Candidates in GOODS-S]
      {First Identification of Direct Collapse Black Hole Candidates in the Early Universe in CANDELS/GOODS-S \vspace{-1.5ex}}
\author[F. Pacucci et al.]
{Fabio Pacucci$^1$ \thanks{fabio.pacucci@sns.it},
Andrea Ferrara$^{1,2}$, Andrea Grazian$^{3}$, Fabrizio Fiore$^{3}$,
\newauthor
Emanuele Giallongo$^{3}$, Simonetta Puccetti$^{4}$ \vspace{0.5ex} \\
$^1$Scuola Normale Superiore, Piazza dei Cavalieri, 7  56126 Pisa, Italy \\
$^2$Kavli Institute for the Physics and Mathematics of the Universe (WPI), Todai Institutes for Advanced Study, \\ 
\, the University of Tokyo 5-1-5 Kashiwanoha, Kashiwa, 277-8583, Japan \\
$^3$INAF-Osservatorio Astronomico di Roma, Via Frascati, 33 - 00078, Monte Porzio Catone, Italy \\
$^4$ASDC-ASI, Via del Politecnico, 00133 Roma, Italy \\
\vspace{-4.0ex}}
         
\date{accepted for publication in MNRAS}
\pubyear{2016}

\begin{document}
\label{firstpage}
\pagerange{\pageref{firstpage}--\pageref{lastpage}}
\maketitle
             
\begin{abstract}
The first black hole seeds, formed when the Universe was younger than $\sim 500 \, \mathrm{Myr}$, are recognized to play an important role for the growth of early ($z \sim 7$) super-massive black holes. While progresses have been made in understanding their formation and growth, their observational signatures remain largely unexplored. As a result, no detection of such sources has been confirmed so far. Supported by numerical simulations, we present a novel photometric method to identify black hole seed candidates in deep multi-wavelength surveys. We predict that these highly-obscured sources are characterized by a steep spectrum in the infrared ($1.6-4.5 \, \mathrm{\mu m}$), i.e. by very red colors. The method selects the only $2$ objects with a robust X-ray detection found in the CANDELS/GOODS-S survey with a photometric redshift $z \gtrsim 6$. Fitting their infrared spectra only with a stellar component would require unrealistic star formation rates ($\gtrsim 2000 \, \mathrm{\Msun \, yr^{-1}}$). To date, the selected objects represent the most promising black hole seed candidates, possibly formed via the direct collapse black hole scenario, with predicted mass $>10^5 \, \mathrm{\Msun}$. While this result is based on the best photometric observations of high-$z$ sources available to date, additional progress is expected from spectroscopic and deeper X-ray data. Upcoming observatories, like the JWST, will greatly expand the scope of this work.
\end{abstract}

\begin{keywords}
quasars: supermassive black holes - black hole physics - galaxies: photometry - cosmology: dark ages, reionization, first stars - cosmology: early Universe - cosmology: observations 
\end{keywords}

\setcounter{footnote}{1}
\newcounter{dummy}
%%%%%%%%%%%%%%%%%%%%%%%%%%%%%%%%%%%%%%%%%%%%%%%%%%%%%%%%%%%%%%%%%%%%%%
%% SECTION 1: INTRODUCTION
%%%%%%%%%%%%%%%%%%%%%%%%%%%%%%%%%%%%%%%%%%%%%%%%%%%%%%%%%%%%%%%%%%%%%%
\section{Introduction}
\label{sec:introduction}
The most distant objects with a spectroscopic redshift measurement detected in the Universe to date are extremely powerful gamma-ray bursts ($z \sim 8.2$, \citealt{Tanvir_2009}), and Lyman-break galaxies ($z \sim 11.1$, \citealt{Oesch_2016}). In the same period of the cosmic time, the reionization epoch, the first Super-Massive Black Holes (SMBHs) were already in place, but the seeds out of which these extremely massive objects were born have not been observed yet.
Several observations (e.g. \citealt{Mortlock_2011, Wu_2015}) have detected the presence of accretion-powered objects with masses in excess of $10^{9-10} \, \mathrm{\Msun}$ already at $z \sim 7$, when the Universe was less than $1$ billion years old. These observations are in tension with the standard theory of black hole growth, which would require, assuming Eddington-limited accretion, a longer time to produce these massive objects \citep{Fan_2006, Haiman_2013} from stellar-mass seeds, born out of the first population of stars (Pop III).
Assuming that low angular momentum gas is always available for feeding, a black hole grows in mass exponentially, with an e-folding time $\sim 0.045 \, \mathrm{Gyr}$, for radiatively efficient accretion models. Starting from a stellar-mass seed ($\lesssim 100 \, \mathrm{\Msun}$), this process would require a constant accretion at the Eddington rate to produce a $\sim 10^9 \, \mathrm{\Msun}$ SMBH by $z \sim 7$, a physically unlikely condition.

Two main solutions to the mass growth enigma have been proposed \citep{Volonteri_2010}. Firstly, the black hole seeds from which the accretion started could have been as massive as $\sim 10^{4-6} \, \mathrm{\Msun}$ \citep{Lodato_Natarajan_2006,Devecchi_2009, Davies_2011}.
One possibility to build up massive black hole seeds at $z \sim 10-15$ is through the Direct Collapse Black Hole (DCBH) scenario \citep{Shang_2010, Johnson_2012}. The collapse of a primordial atomic-cooling halo (with a virial temperature $T_{\mathrm{vir}} \gtrsim 10^4 \, \mathrm{K}$) may lead, in the presence of a strong flux of Lyman-Werner photons (energy $h\nu =11.2-13.6 \, \mathrm{eV}$) dissociating the $\mathrm{H}_2$ and thus preventing gas fragmentation, to the formation of DCBHs with a typical mass around $10^5 \, \mathrm{\Msun}$ \citep{Ferrara_2014}.
Secondly, the growth rates might not be capped by the Eddington value \citep{Volonteri_2005, Alexander_2014, Madau_2014, Volonteri_2014, PVF_2015} $\dot{M}_{\mathrm{Edd}} \equiv L_{\mathrm{Edd}}/(\eta c^2)$, where $L_{\mathrm{Edd}} \approx 1.2 \times 10^{38} M_{\bullet} \, \mathrm{erg \, s^{-1}}$, $M_{\bullet}$ is the black hole mass in solar masses, $\eta$ is the efficiency factor for mass-energy conversion and $c$ is the speed of light.
In highly-obscured environments, radiation trapping could allow super-Eddington accretion rates to take place, dramatically speeding up the black hole growth.

At the present time no detection of early SMBH progenitors has been confirmed. This may be due to their extreme faintness, but possibly also to the current lack of clear theoretical indications about their mass, host halo properties and typical accretion rates, resulting in large uncertainties in the prediction of their observational signatures.
We proposed (\citealt{Pallottini_Pacucci_2015_CR7}, see also \citealt{Agarwal_2015,Hartwig_2015, Visbal_2016, Smith_2016, Smidt_2016, Dijkstra_2016_CR7}) that a $z \approx 6.6$ object named CR7, the brightest Ly-$\alpha$ emitter discovered so far \citep{Sobral_2015}, could be powered by a DCBH with an initial mass $\sim 10^5 \, \mathrm{M_{\odot}}$, associated with a standard stellar emission. This was motivated by the peculiarities of its spectrum, namely: (i) very strong Ly-$\alpha$ and He II lines emission, and (ii) absence of metal lines, within the detection threshold. This proposition needs deeper X-ray observations to be confirmed or rejected.

In previous works \citep{Pacucci_2015, PFVD_2015}, employing a combination of radiation-hydrodynamic simulations and spectral synthesis codes, we computed the time-evolving spectrum emerging from the halo hosting a black hole seed.
The bulk of the emission occurs in the observed infrared-submm ($1-1000 \, \mathrm{\mu m}$) and X-ray ($0.1 - 100 \, \mathrm{keV}$) bands. Here we present a method to select DCBH candidates in ultra-deep fields purely based on infrared photometry. 

The outline of this paper is as follows. In $\S 2$ we describe the numerical setup of our simulations, while $\S 3$ presents the photometric tool that we developed. In $\S 4$ we show our results, including the identification of two DCBH candidates in CANDELS/GOODS-S. Finally, in $\S 5$ we discuss the caveats of this work and in $\S 6$ we provide some further discussion and the conclusions.
Throughout, we adopt recent Planck cosmological parameters \citep{Planck_2015}: $(\Omega_m, \Omega_{\Lambda}, \Omega_b, h, n_s, \sigma_8 )= (0.32, 0.68, 0.05, 0.67, 0.96, 0.83)$.

%%%%%%%%%%%%%%%%%%%%%%%%%%%%%%%%%%%%%%%%%%%%%%%%%%%%%%%%%%%%%%%%%%%%%%
%% SECTION 2: NUMERICAL SETUP
%%%%%%%%%%%%%%%%%%%%%%%%%%%%%%%%%%%%%%%%%%%%%%%%%%%%%%%%%%%%%%%%%%%%%%
\section{Numerical Implementation}
\label{sec:implementation}
In this Section we introduce the physical and numerical implementation of our simulations.
The interested reader is referred to \cite{Pacucci_2015} (radiation-hydrodynamic module) and to \cite{PFVD_2015} (spectral module) for a much more detailed description.

\subsection{General physical framework}
A high-$z$ black hole seed, with initial mass $M_{\bullet}(t=0)$ in the range $10^{4-5} \, \mathrm{\Msun}$, is placed at the center of a dark matter halo with primordial composition gas (H and He, with helium fraction $Y_P \approx 0.247$, \citealt{Planck_2015}), total mass (baryonic and dark matter) $M_h$ and virial temperature $T_{\mathrm{vir}} = T_{\mathrm{vir}}(M_h,z)  \sim 10^4 \, \mathrm{K}$ \citep{BL01}:
\begin{equation}
T_{\rm vir} \approx 1.98\times 10^4\ \left(\frac{\mu}{0.6}\right)
\left(\frac{M_h}{10^8\ h^{-1} \ \mathrm{\Msun} }\right)^{2/3} 
\left(\frac{1+z}{10}\right)\ {\rm K} \, .
\label{T_vir_eq}
\end{equation}
In this formula, $\mu$ is the mean molecular weight and $h$ is the reduced Hubble constant.
At $z \sim 10$, $T_{\mathrm{vir}} \sim 10^4 \, \mathrm{K}$ corresponds to $M_h \sim 10^8 \, \mathrm{\Msun}$. 
In this work, we do not assume any specific relation between the initial black hole mass $M_{\bullet}(t=0)$ and its host halo mass $M_h$.  The mass of a black hole seed is assigned following the initial mass function derived in \cite{Ferrara_2014}.
The black hole accretes mass from the inner regions of the host halo, with an accretion rate, $\dot{M}$, self-regulated by the combined effects of gravity, gas pressure and radiation pressure.

\subsection{Radiation-hydrodynamics}
\label{subsec:rad_hydro}
Our radiation-hydrodynamic code \citep{Pacucci_2015} evolves self-consistently the standard system of ideal, non-relativistic Euler's equations for a gas accreting radially onto the central black hole, assumed at rest and already formed at $t=0$.
We do not simulate the formation process of the massive black hole, but our implicit assumption is that it forms via the DCBH scenario, since: (i) the host halo is metal-free, and (ii) the surrounding gas has not been photo-evaporated by Pop III stellar emission. Other formation channels would not comply with these two assumptions.

The spatial domain of the simulations spans from $0.002 \, \mathrm{pc}$ to $200 \, \mathrm{pc}$, largely encompassing the characteristic spatial scale for accretion, the Bondi radius, which, for a $10^5 \, \mathrm{\Msun}$ black hole, is $\sim 3 \, \mathrm{pc}$.
The gas initially follows the isothermal density profile derived from the simulations in \cite{Latif_2013}, where $\rho(r) \propto (r/a)^{-2}$ and $a \sim 2 \, \mathrm{pc}$ is the core radius of the baryonic matter distribution. 
This density profile is the one resulting after the formation of a massive seed of mass $\sim 10^5 \, \mathrm{\Msun}$ at the halo center \citep{Latif_2013c, Latif_2014b}.

The value of the black hole mass $M_{\bullet}(t)$ changes with time, due to the accretion, modeled with the accretion rate $\dot{M}_{\bullet}=4 \pi r^2 \rho |v|$ ($r$ is the radial coordinate, $\rho$ the gas density and $v$ is the radial velocity, all computed at the inner boundary of the spatial domain).
The accretion rate generates an emitted bolometric luminosity $L_{\rm bol}$ computed as:
\begin{equation}
L_{\rm bol} \equiv \eta c^2 \dot{M} \, ,
\label{l_bol}
\end{equation}
where $\eta \approx 0.1$.
The radiation pressure accelerates the gas via $a_{rad}(r) = \kappa(\rho, T) L_{\rm bol}(r)/(4 \pi r^2 c)$, where the gas opacity $\kappa(\rho, T)$, function of the gas density and temperature, includes Thomson \citep{Begelman_2008} and bound-free terms.

\subsection{Observed DCBH spectrum}
\label{subsec:spectrum}
Our radiation-hydrodynamic code computes the frequency-integrated radiative transfer through the gas, taking into consideration the appropriate: (i) cooling and heating terms, (ii) matter-to-radiation coupling, and (iii) energy propagation through a two-stream approximation method. Full details are provided in \cite{Pacucci_2015}.
The {\em frequency-dependent} radiative transfer through the host halo is then performed using the spectral synthesis code \texttt{CLOUDY} \citep{Cloudy}. This code computes the time-evolving spectrum emerging from the host halo \citep{PFVD_2015} employing as input data: (i) the spatial profiles for hydrogen number density and temperature, (ii) the source spectrum of the central object, and (iii) the bolometric luminosity of the source (computed self-consistently with Eq. \ref{l_bol}).
The source spectral energy distribution (SED), extended from far-infrared to hard X-ray, is a standard Active Galactic Nuclei (AGN) spectrum, computed for a metal-free gas \citep{Yue_2013}. The source spectrum depends on the black hole mass, and therefore evolves with time as $M_{\bullet}$ increases.

%%%%%%%%%%%%%%%%%%%%%%%%%%%%%%%%%%%%%%%%%%%%%%%%%%%%%%%%%%%%%%%%%%%%%%
%% SECTION 3: PHOTOMETRY AND X-RAYS
%%%%%%%%%%%%%%%%%%%%%%%%%%%%%%%%%%%%%%%%%%%%%%%%%%%%%%%%%%%%%%%%%%%%%%
\section{Photometry and SED Fitting}
\label{sec:photometry}
In this Section we describe the main photometric tool used in this work, along with the optical/infrared and X-ray data employed.

\subsection{The color-color plot}
\label{subsec:color_color}
The shape of the spectrum emerging from a high-$z$ dark matter halo hosting a DCBH depends on the black hole mass and on the absorbing hydrogen column density of the surrounding gas. If the halo is Compton-thick ($N_{H} \gtrsim 1.5 \times 10^{24} \, \mathrm{cm^{-2}}$) high-energy photons ($\lesssim 10-50 \, \mathrm{keV}$) are largely absorbed and the DCBH may be invisible in the X-ray. Nonetheless, sufficiently massive black hole seeds, $M_{\bullet}(t=0) \gtrsim 10^{4-5} \, \mathrm{\Msun}$, are always visible in the infrared \citep{PFVD_2015}, due to the reprocessing of high-energy radiation into lower-energy photons. Moreover, the infrared spectrum is less dependent on the specifics of the accretion flow.

For these reasons, we devised a method to select DCBH candidates through infrared photometry.
The computation of the predicted photometry for a DCBH follows from: (i) the time evolution of the spectrum emerging from the host halo, and (ii) the shapes of the photometric filters employed.
Denoting by ${\cal F}(F)$ the flux measured through the filter $F$, the generic color index for the couple of filters $A$ and $B$ is defined as follows: $A-B \equiv -2.5 \, \mathrm{Log}_{10} \left[ {\cal F}(A)/{\cal F}(B) \right]$.

\begin{figure}
\vspace{-1\baselineskip}
\hspace{-0.5cm}
\begin{center}
\includegraphics[angle=0,width=0.5\textwidth]{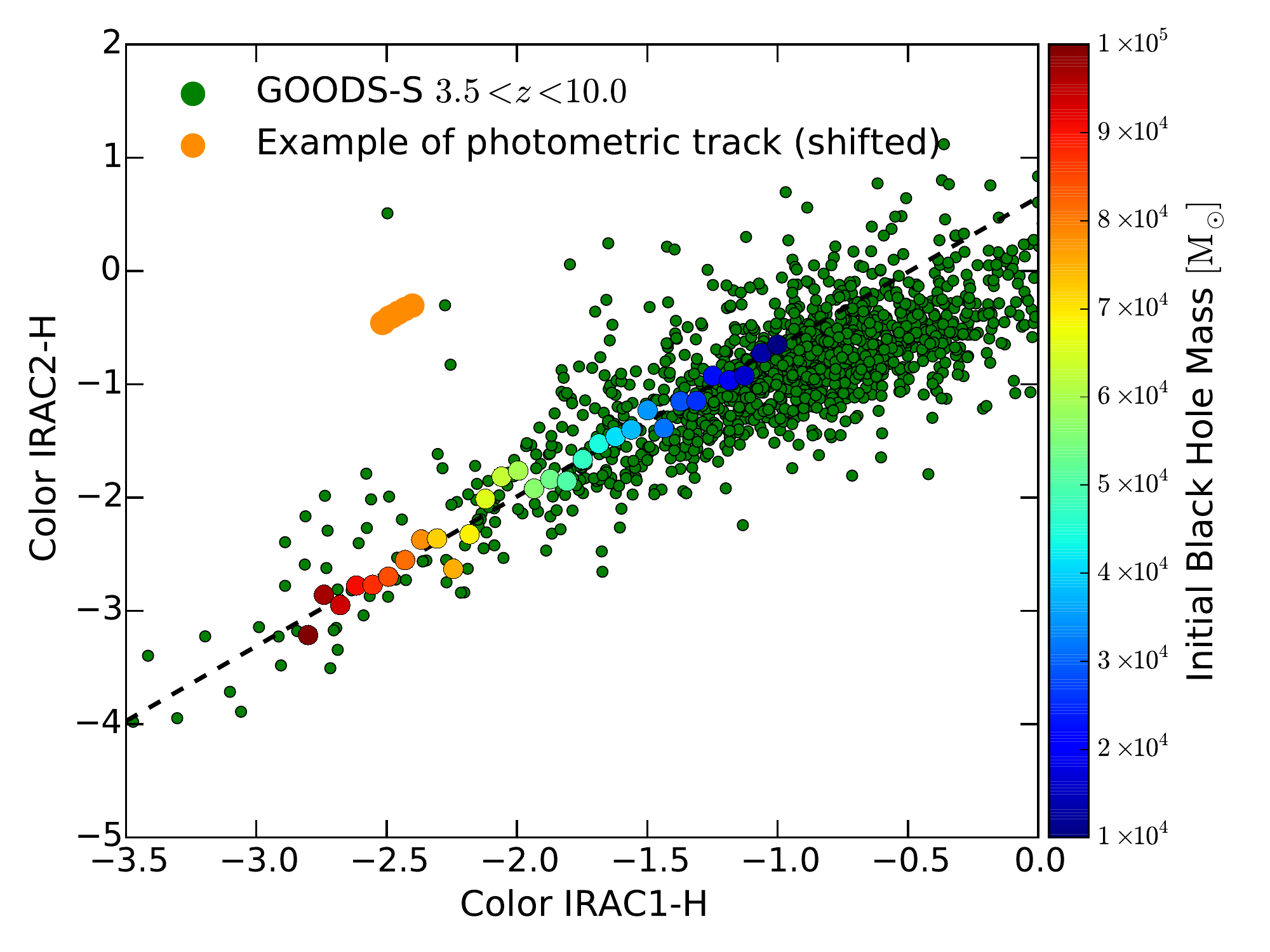}
\caption{Color-color diagram for the infrared filters H, IRAC1 and IRAC2. GOODS-S objects, brighter than the 27th magnitude in the H band ($\mathrm{H<27}$) and with $3.5 \lesssim z \lesssim 10$, are shown with green points. Numerical predictions for the colors of DCBHs are shown, at $z \sim 7$, with filled circles, whose color depends on the initial mass of the seed (see the color-bar). Larger black hole masses are associated with redder spectra (i.e. more negative colors). All colors are \textit{observed quantities}. An example of a photometric track (see Section \ref{subsec:color_evolution_column_density}) for a DCBH of initial mass $\sim 8 \times 10^{4} \, \mathrm{\Msun}$ is shown in orange. Its position has been shifted vertically to avoid information overload.}
\label{fig:colors_total}
\end{center}
\end{figure}

We employ three photometric filters to probe the infrared spectrum: the HST H ($1.6 \, \mathrm{\mu m}$), the Spitzer IRAC1 ($3.6 \, \mathrm{\mu m}$) and IRAC2 ($4.5 \, \mathrm{\mu m}$).
We construct a color-color diagram (Fig. \ref{fig:colors_total}) in which we plot both data points for objects in the CANDELS/GOODS-S multi-wavelength survey (see Section \ref{subsec:candels} for a description of the data) and our predictions for the colors of DCBHs. All colors are \textit{observed quantities}.
The data points roughly fall on a line since the observed infrared spectra are to a good approximation power laws, with a broad range of slopes. The position in the color-color plot of DCBH seeds with initial masses in the range $10^{4-5} \, \mathrm{\Msun}$, computed via radiation-hydrodynamic simulations, is shown with filled circles whose color depends on $M_{\bullet}(t=0)$, as shown in the color-bar.

To compute the photometric points for DCBHs as a function of their mass, a total number of $\sim 100$ simulations have been performed, initializing the black hole seed with masses in the range $10^{4-5} \, \mathrm{\Msun}$. For each simulation, the mass of the host halo has been modified in order to set the virial temperature $T_{\mathrm{vir}} \sim 10^4 \, \mathrm{K}$ for the redshift of choice (Eq. \ref{T_vir_eq}). Different sets of simulations have been performed at redshifts $z=13$, $z=10$ and $z=7$. Since the shape of the SED at the wavelengths of our interest is roughly a power law, the effect of a redshift change on the position of the photometric points is small. 
Each simulation has been run for a time which varies with $M_h$ and with $M_{\bullet}(t=0)$, between $\sim10 \, \mathrm{Myr}$ and $\sim 80 \, \mathrm{Myr}$, in order to reach a steady state accretion without depleting the gas reservoir. Sections \ref{subsec:color_evolution_column_density} and \ref{subsec:color_evolution_mass} describe the dependence of DCBH colors on column density and on black hole mass.

Our method is perfectly adaptable to the JWST photometric system.
The photometric filters F150W (center wavelength $\lambda_0=1.5 \, \mathrm{\mu m}$), F356W ($\lambda_0=3.6 \, \mathrm{\mu m}$) and F444W ($\lambda_0=4.4 \, \mathrm{\mu m}$) are equivalents to the H, IRAC1 and IRAC2 filters, respectively, and may be used to implement our selection method with the JWST.

\subsection{The CANDELS/GOODS-S survey}
\label{subsec:candels}
The photometric data employed in this work are based on the GOODS-S field of the official CANDELS catalogues \citep{Guo_2013}. Object selection has been performed in the H band of the near-infrared Wide Field Camera-3.
The covering area of the GOODS-S survey is $\sim 170 \, \mathrm{arcmin}^2$, to a mean $5\sigma$ depth of $27.5$ magnitudes in the H band.
These imaging data include photometry over a wide range of wavelengths, from the U band ($0.36 \, \mathrm{\mu m}$) to the IRAC4 band ($7.9 \, \mathrm{\mu m}$). Importantly for our purposes, the catalog includes very deep imaging with the IRAC instrument from the Spitzer Extended Deep Survey \citep{Ashby_2013}, covering the CANDELS fields to a $3\sigma$ depth of $26$ AB magnitude at both $3.6 \, \mathrm{\mu m}$ (IRAC1) and $4.5 \, \mathrm{\mu m}$ (IRAC2).
Overall, the detection of objects in the GOODS-S field is realistic up to $27.5-28.0$ magnitudes in the H band, at $\sim 90\%$ completeness limit \citep{Grazian_2015}.

For the photometric data that we employed in this work, each source has been visually inspected.
The total number of sources detected in the GOODS-S field is $34930$.
In our redshift range of interest, $3.5 \lesssim z \lesssim 10$ (see Fig. \ref{fig:colors_total}), the total number of objects is $2037$, while at high redshifts ($6 \lesssim z \lesssim 10$) is $97$.
In our analysis we included only GOODS-S objects with a precise value (i.e. not upper limits) for the three magnitudes H, IRAC1 and IRAC2.
In total, there are $2272$ spectroscopic redshifts of good quality in the GOODS-S field. The number of spectroscopic redshifts available at $z>3.5$ is $173$ ($25$ at $z>6.5$).
For sources lacking spectroscopic information, photometric redshifts were computed by optimally-combining six different photometric redshifts \citep{Dahlen_2013}. For a broader discussion on the uncertainties in the redshift determination, see Section \ref{subsec:caveats_red}.

\subsection{X-ray detected objects in the GOODS-S survey}
\label{subsec:x_ray_candels}
\begin{figure*}
\vspace{-1\baselineskip}
\hspace{-0.5cm}
\includegraphics[angle=0,width=0.29\textwidth]{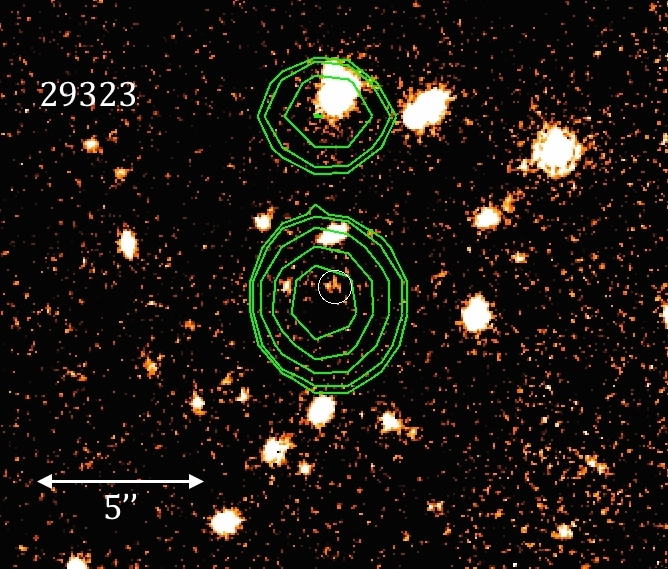}
\includegraphics[angle=0,width=0.29\textwidth]{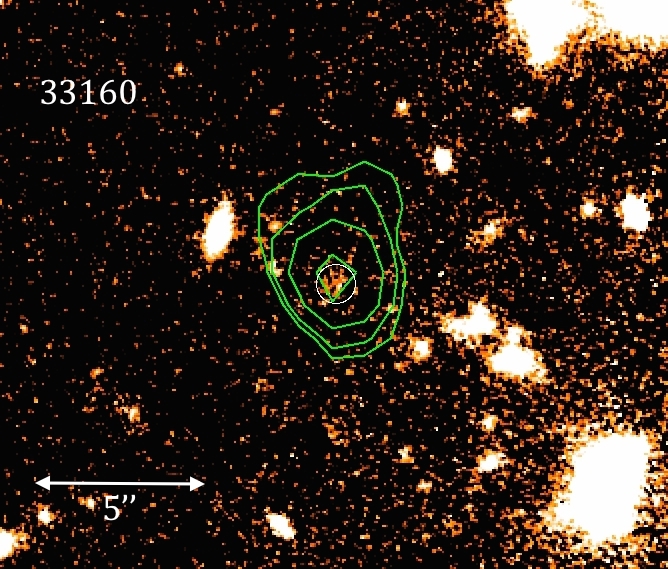}
\includegraphics[angle=0,width=0.40\textwidth]{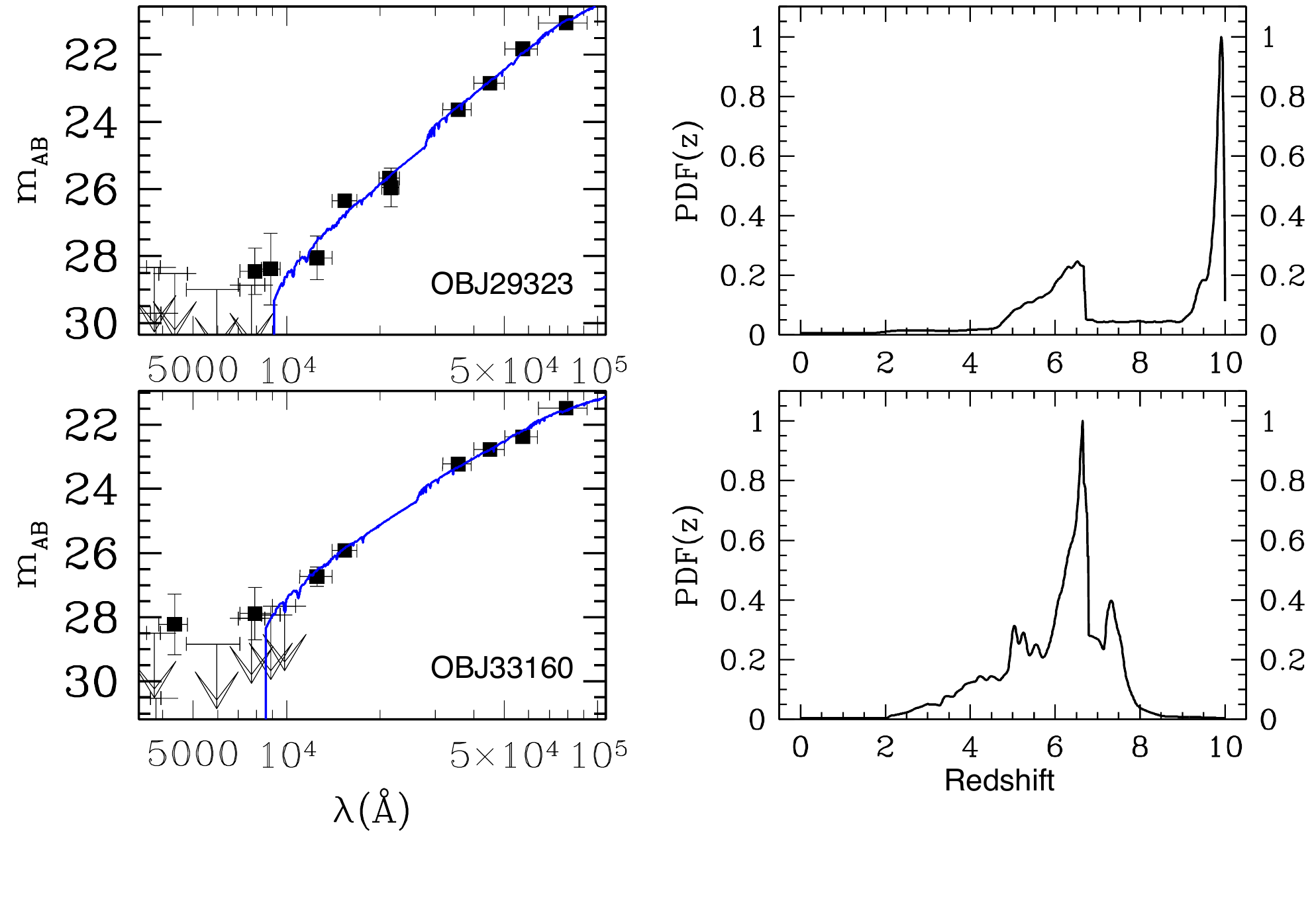}\\
\vspace{-0.1cm}
\caption{H band images with X-ray contours ($1-4 \, \mathrm{keV}$) of the two GOODS-S sources selected as DCBH candidates. X-ray contours are from the Chandra $7 \, \mathrm{Ms}$ field, in square root scale, from $2\sigma$ to $10\sigma$. H band counterparts are circled. On the right, probability distribution functions (taken from \citealt{Giallongo_2015}) for the photometric redshifts of the same objects, along with their optical/infrared spectra, fitted from their photometry.}
\label{fig:pdf}
\end{figure*}

In a recent work \citep{Giallongo_2015}, $22$ faint AGN candidates were identified in the GOODS-S field as having an X-ray counterpart (but see \citealt{Weigel_2015} and \citealt{Cappelluti_2015}).
The selection method is the following one: high-$z$ galaxies are selected in the H band down to very faint levels ($\mathrm{H} \leq 27$) using reliable photometric redshifts. At $z>4$, this corresponds to a selection criterion based on the galaxy rest-frame UV flux. AGN candidates are picked up from this parent sample if they show X-ray fluxes above a threshold of $F_X \sim 1.5 \times 10^{-17} \, \mathrm{erg\, cm^{-2} \, s^{-1}}$ in the soft energy band ($0.5-2\, \mathrm{keV}$).

For $3$ of these sources the photometric redshift is $z\gtrsim 6$: object $33160$ at $z \approx 6.06$, object $29323$ at $z \approx 9.73$ and object $28476$ at $z \approx 6.26$. H band images of the first two sources, with X-ray contours from the Chandra $7 \, \mathrm{Ms}$ field, are shown in Fig. \ref{fig:pdf}. As we will show in Section \ref{sec:results}, these two sources are our DCBH candidates. The photometric redshift of $29323$ is more uncertain, but its probability distribution function suggests that $z \gtrsim 6$ with high confidence (see Fig. \ref{fig:pdf}). 
Objects $33160$ and $29323$, both at $z\gtrsim 6$, are to be considered robust detections, being also included in the \cite{Xue_2011} catalog, with identification numbers 85 and 156, respectively. Moreover, the source 29323 is also detected in \cite{Cappelluti_2015}, but these authors suggest that the determination of its photometric redshift could be affected by potential artifacts in the SED.

Instead, object $28476$ is a possible X-ray detection: for this source there are $15$ detected counts in the soft-band ($0.5-2 \, \mathrm{keV}$), compared to $\sim 5$ background counts within a radius of 3 pixel, corresponding to a probability of spurious detection of $\sim 10^{-4}$. Anyway, the association between this X-ray source and its optical counterpart is more problematic. A full description, including coordinates, of all the 22 X-ray detected sources is provided in \cite{Giallongo_2015}.

%%%%%%%%%%%%%%%%%%%%%%%%%%%%%%%%%%%%%%%%%%%%%%%%%%%%%%%%%%%%%%%%%%%%%%
%% SECTION 4: RESULTS
%%%%%%%%%%%%%%%%%%%%%%%%%%%%%%%%%%%%%%%%%%%%%%%%%%%%%%%%%%%%%%%%%%%%%%
\section{Results}
\label{sec:results}
In this Section we describe how the photometry of DCBHs changes with the column density of the host halo and with the black hole mass. Then, we present the black hole seed candidates selected in the CANDELS/GOODS-S field.

\subsection{Column density dependence of DCBH colors}
\label{subsec:color_evolution_column_density}
The initial black hole mass $M_{\bullet}(t=0)$ is set for any given simulation in the range $10^{4-5} \, \mathrm{\Msun}$. At some redshift $z$, the gas mass $M_g$ of the host halo is also set from Eq. \ref{T_vir_eq}, which in turn translates into an initial column density $N_H$, with values generally in the Compton-thick range. The column density controls how photons with different frequencies are absorbed or transmitted. 
As long as the column density is large ($N_H \gtrsim 1.5 \times 10^{24} \, \mathrm{cm^{-2}}$), the photometry of DCBHs does not vary largely. As the black hole grows with time, its bolometric luminosity increases as well, but this effect does not modify the colors, since the increase in flux is similar in each photometric band. Once the column density falls below the Compton-thick limit, the evolution is very rapid and short-lived ($\lesssim 1-2 \, \mathrm{Myr}$, \citealt{Pacucci_2015}): the DCBH performs a short photometric track in the color-color plot (see Fig. \ref{fig:colors_total}).
The filled circles shown in Fig. \ref{fig:colors_total} are computed at $z=7$ and are averages, weighted over time, of the photometric track performed in each simulation.

\subsection{Mass dependence of DCBH colors}
\label{subsec:color_evolution_mass}
The trend of DCBH colors with the black hole mass (Fig. \ref{fig:colors_total}) may be simply explained with the following argument.
The column density $N_H$ is proportional to the gas mass $M_g$ and to the radius of the baryonic matter distribution $R_g$ in the following way: $N_H \propto M_g/R_g^2$. Defining the concentration parameter of the gas distribution as ${\cal C} \equiv M_g/R_g \propto M_{\bullet}$ (larger black holes are more efficient and rapid in concentrating the gas around them), we derive that the column density of halos hosting more massive black holes is larger: $N_H \propto M_g/R_g^2 \propto {\cal C}/R_g \propto M_{\bullet}$. Here we are assuming that, on sufficiently large spatial scales ($\gg R_B$), the characteristic radius of the baryonic matter distribution of the host halo is not modified by the mass of the black hole seed that forms at its center.
The gas around larger black holes is more concentrated and then more efficient in reprocessing the radiation to lower energies.
Consequently, the steepness of the infrared spectrum of massive black hole seeds is directly proportional to the black hole mass.

\subsection{Identification of DCBH candidates in GOODS-S}
\label{subsec:candidates}
\begin{figure}
\vspace{-1\baselineskip}% \hspace{-0.5cm}
\begin{center}
\includegraphics[angle=0,width=0.5\textwidth]{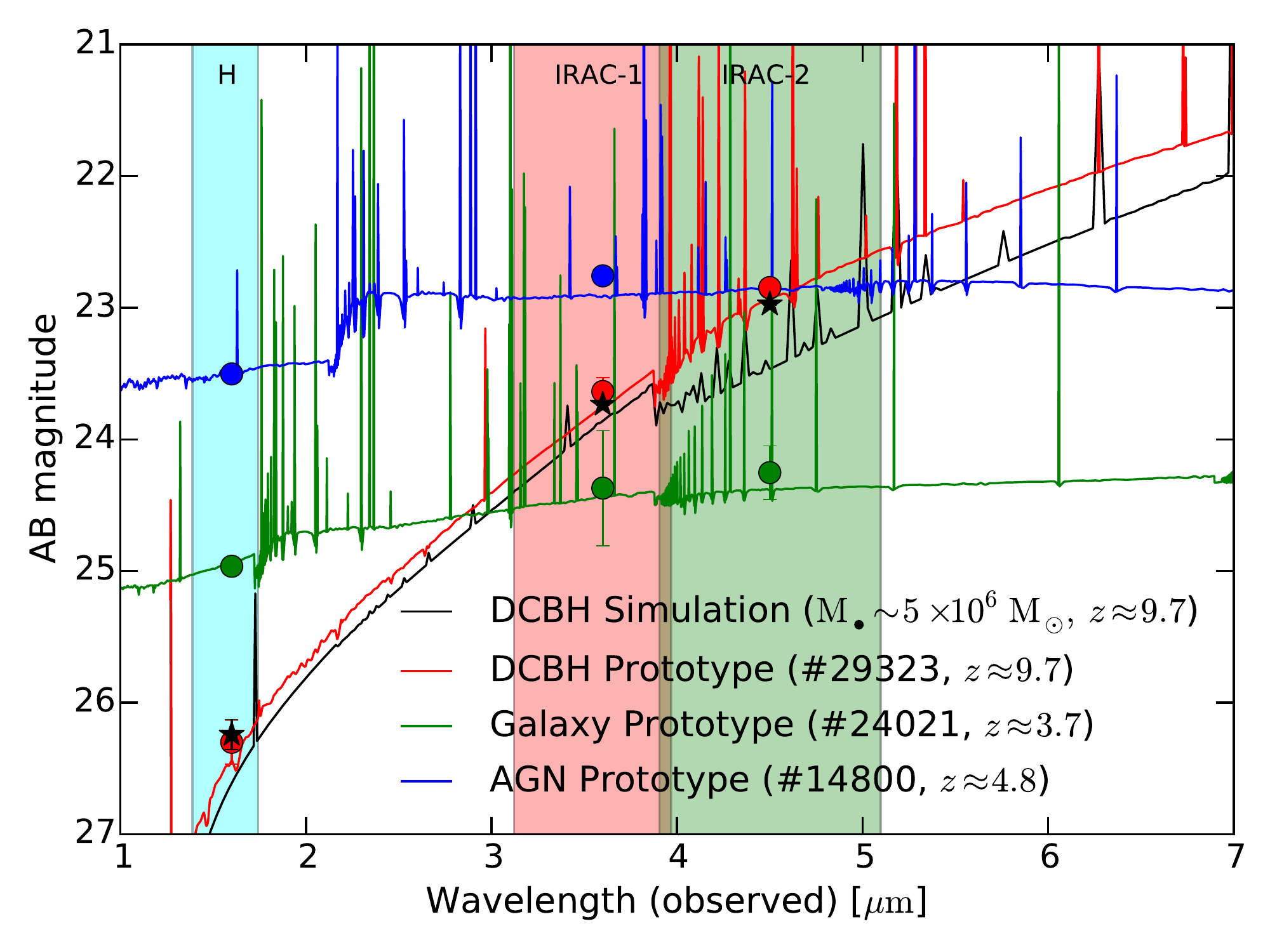}
\caption{Comparison between the stellar SEDs of three GOODS-S objects with the computed SED of a $\sim 5 \times 10^6 \, \mathrm{\Msun}$ black hole, born out of a DCBH with initial mass around $\sim 10^5 \, \mathrm{\Msun}$. The three photometric bands employed in our work are shown as shaded regions. Filled circles (observations) and stars (numerical simulations) show the magnitudes in the three filters, with error bars. Objects $29323$ and $14800$ have X-ray counterparts (i.e. they are likely associated with a black hole), while $24021$ has not (i.e. it is likely a normal galaxy). Moreover, object $29323$ is characterized by very negative colors (i.e. its infrared SED is very steep, as we predict for DCBHs), while objects $14800$ and $24021$ are not. The steepness of the SED and the infrared magnitudes for the object $29323$ are well fitted by the spectrum predicted for a $\sim 5 \times 10^6 \, \mathrm{\Msun}$ black hole. In the computed SED for a DCBH, the He II line ($0.164 \, \mathrm{\mu m}$ rest-frame) is visible and it is marginally inside the H band at $z \approx 9.7$.}
\label{fig:seds}
\end{center}
\end{figure}

DCBHs with larger masses are associated with steeper (i.e. redder) infrared spectra. These massive objects extend down to a region in the color-color diagram where there is a relative paucity of GOODS-S objects (see Fig. \ref{fig:colors_total}).
We pick three GOODS-S objects characterized by different observational features and compare in Fig. \ref{fig:seds} their SEDs with the one computed for a $\sim 5 \times 10^6 \, \mathrm{\Msun}$ black hole, grown from a DCBH with initial mass around $\sim 10^5 \, \mathrm{\Msun}$. The SED fitting has been executed with a standard stellar population model \citep{Bruzual_Charlot_2003}. As described in \cite{Castellano_2014} and in \cite{Grazian_2015}, we have included the contribution from nebular emission lines following \cite{Schaerer_deBarros_2009}. 

Objects $29323$ (red line) and $14800$ (blue line) are objects with an X-ray counterpart whose colors are very negative ($29323$) or close to zero ($14800$). Object $24021$ (green line) is instead an object without an observed X-ray emission, whose colors are close to zero.
The predicted infrared SED for a DCBH is redder than the ones observed for normal galaxies (e.g. $24021$, in which radiation comes predominantly from stars) and AGNs (e.g. $14800$, in which accretion-powered radiation co-exists with the stellar one). Current theoretical models suggest that most of the high-$z$ black hole growth occurred into heavily-obscured hosts (see e.g. \citealt{Comastri_2015}), providing a theoretical ground for the redness of the infrared SED of DCBHs.

Objects in the GOODS-S field may be detected up to $\mathrm{H \approx 28}$ with a completeness $\sim 90\%$. From our simulations, we conclude that DCBH seeds with initial masses below $\sim 6 \times 10^4 \, \mathrm{\Msun}$ cannot be detected in current surveys, assuming realistic growth rates.
Then, we predict to observe DCBH candidates with initial masses $\gtrsim 6 \times 10^4 \, \mathrm{\Msun}$, i.e. with colors $\mathrm{IRAC1-H} \lesssim -1.8$ and $\mathrm{IRAC2-H} \lesssim -1.8$ (see Fig. \ref{fig:colors_total}). The infrared SEDs of objects within this area of the color-color diagram are very red: they are promising DCBH candidates.

\begin{figure}
\vspace{-1\baselineskip}
\hspace{-0.5cm}
\begin{center}
\includegraphics[angle=0,width=0.5\textwidth]{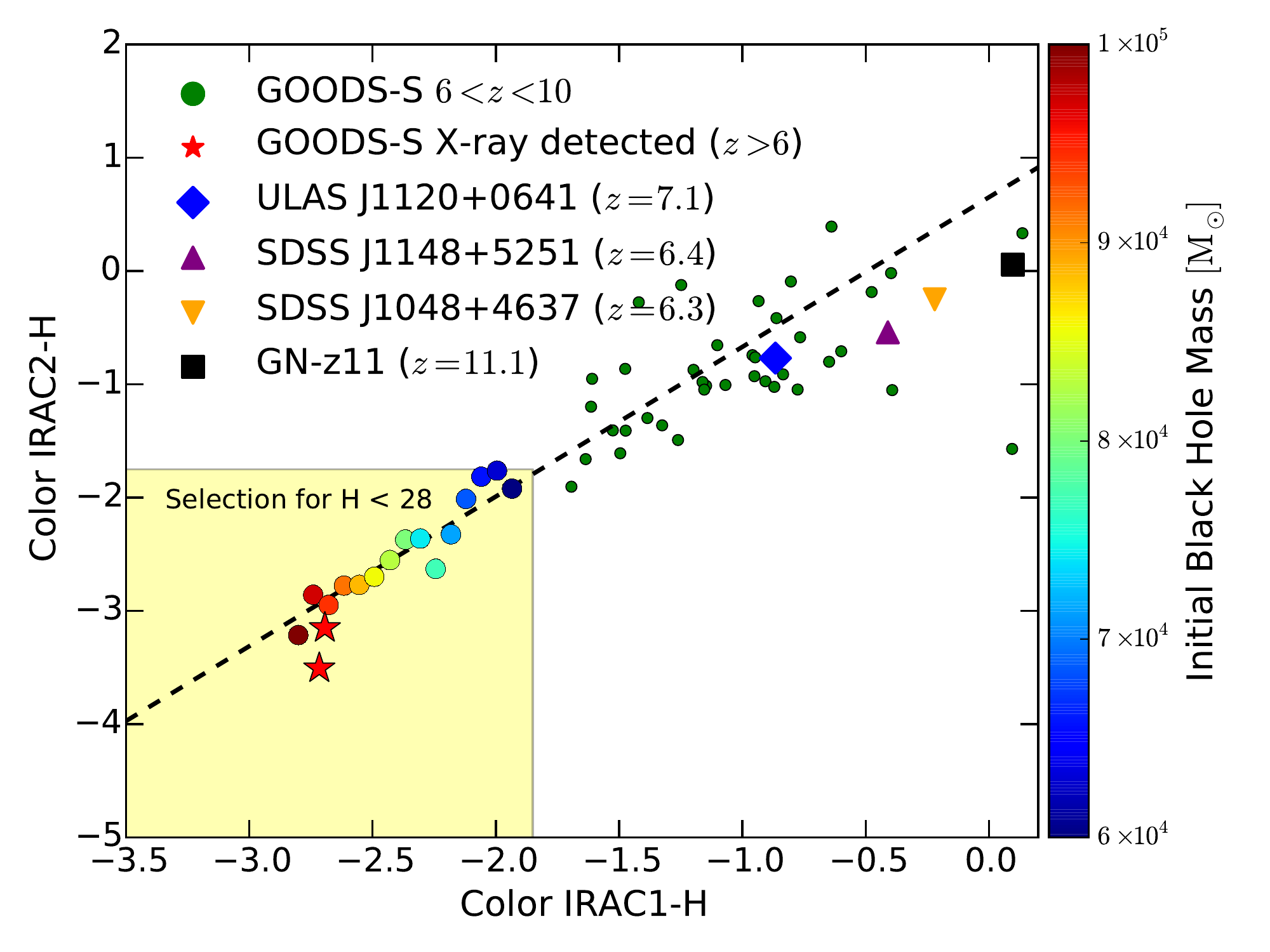}
\caption{Color-color diagram for objects at $z \gtrsim 6$. The yellow-shaded area indicates the region in the diagram where we expect SEDs compatible with the ones predicted for DCBHs, and observable with current surveys. Green points are GOODS-S objects without an X-ray counterpart, while red stars are GOODS-S objects detected in the X-ray and likely powered by accretion onto a collapsed object. Filled circles are, as in Fig. \ref{fig:colors_total}, DCBH simulations. In addition, we report the position in the color-color plot of several $z>6$ sources. ULAS J1120+0641 \citep{Mortlock_2011,Barnett_2015} is the most distant known QSO ($z \approx 7.1$). SDSS J1148+5251 \citep{Fan_2003} and SDSS J1048+4637 \citep{Maiolino_2004} are two highly-obscured and dusty QSOs at $z\approx 6.4$ and $z \approx 6.3$, respectively. GN-z11 \citep{Oesch_2016} is the highest redshift galaxy discovered to date ($z \approx 11.1$).}
\label{fig:colors_highz}
\end{center}
\end{figure}

The two GOODS-S objects with a robust X-ray detection at $z \gtrsim 6$ (see Section \ref{subsec:x_ray_candels}) fall within our selection region for DCBH candidates, as shown in Fig. \ref{fig:colors_highz}, which is the main result of this work.
At $z \gtrsim 6$ the photometry of these sources is very similar to the one predicted for DCBHs, characterized by a very red infrared SED, caused by large absorbing column densities.
At lower redshifts ($z \lesssim 6$) these pristine spectra are likely modified by several events, like star formation and large-scale outflows, which decrease the column density of the host halo. Moreover, also the pollution of the gas by metals may contribute to the modification of the SED.
In Section \ref{subsec:SFR} we discuss other kind of sources possibly mimicking DCBH colors.
Conservatively, we have not included in Fig. \ref{fig:colors_highz} the position of the $z \gtrsim 6$ object $28476$, which also falls in our selection region, but whose X-ray detection is less robust.

In Fig. \ref{fig:colors_highz} we also show the position of other high-$z$ objects, to compare it with the locations of our DCBH candidates. ULAS J1120+0641 \citep{Mortlock_2011,Barnett_2015} is the most distant known Quasi-Stellar Object (QSO, $z \approx 7.1$), powered by a SMBH with an estimated mass $\sim 10^9 \, \mathrm{\Msun}$. SDSS J1148+5251 \citep{Fan_2003} and SDSS J1048+4637 \citep{Maiolino_2004} are two highly-obscured and dusty QSOs at $z\approx 6.4$ and $z \approx 6.3$, respectively. GN-z11 \citep{Oesch_2016} is the highest redshift galaxy discovered to date ($z \approx 11.1$).
The position of these objects in the color-color diagram is outside the region of detectability for sources powered by a DCBH: their H band magnitudes would be too faint to be currently observed. Consequently, our photometric selection method seems to separate the spectra of pristine DCBHs from those of other contaminant high-$z$ sources.
 
For the couple of $z \gtrsim 6$ X-ray detected sources in the CANDELS/GOODS-S field we estimate a black hole mass in excess of $\sim 10^{6} \, \mathrm{\Msun}$.
For the source $33160$ we are able to reproduce simultaneously with simulations both its H band magnitude, $\mathrm{H=25.9}$, and its $2-10 \, \mathrm{keV}$ X-ray luminosity emerging from the host halo, $\mathrm{Log}_{10}(L_x) = 43.65$. Since we assume the photometric redshift reported in \cite{Giallongo_2015}, also its X-ray flux is automatically reproduced. Source $29323$ at $z \approx 9.73$ has a X-ray luminosity slightly above our predicted range (corresponding to a maximum black hole mass of $\sim 5 \times 10^6 \, \mathrm{\Msun}$), while we successfully reproduce its H band magnitude (see Fig. \ref{fig:seds}). If this object is instead at redshift $z \lesssim 6$, its X-ray luminosity would be downgraded to $\mathrm{Log}_{10}(L_x) \lesssim 43.5$, well inside our predicted range.
More generally, the X-ray detected objects in CANDELS/GOODS-S that are selected by our method (at any redshift) are compatible with the X-ray luminosities ($2-10 \, \mathrm{keV}$) predicted for a population of DCBHs with initial masses in the range $6\times 10^4 - 10^5 \, \mathrm{\Msun}$. Within this ensemble of sources, more negative colors are associated with a larger X-ray luminosity (i.e. with a larger black hole mass, as in Fig. \ref{fig:colors_total}). 

In the simulated DCBH spectrum of the object 29323 in Fig. \ref{fig:seds}, the He II line ($0.164 \, \mathrm{\mu m}$ rest-frame) is visible and it is marginally inside the H band at $z \approx 9.7$. This emission line is an important indicator of DCBH activity (see e.g. \citealt{Pallottini_Pacucci_2015_CR7}) and its detection, with future spectroscopic observations, may be a discriminant in the determination of the real nature of this high-$z$ source.

\subsection{Star formation rates in the X-ray detected sample}
\label{subsec:SFR}
The photometry of all the X-ray detected objects in GOODS-S has been fitted by a stellar SED fitting model \citep{Bruzual_Charlot_2003}. The SED fitting assumes that all the luminosity is produced by stars, without any contribution from accreting objects.

\begin{figure}
\vspace{-1\baselineskip}
\hspace{-0.5cm}
\begin{center}
\includegraphics[angle=0,width=0.5\textwidth]{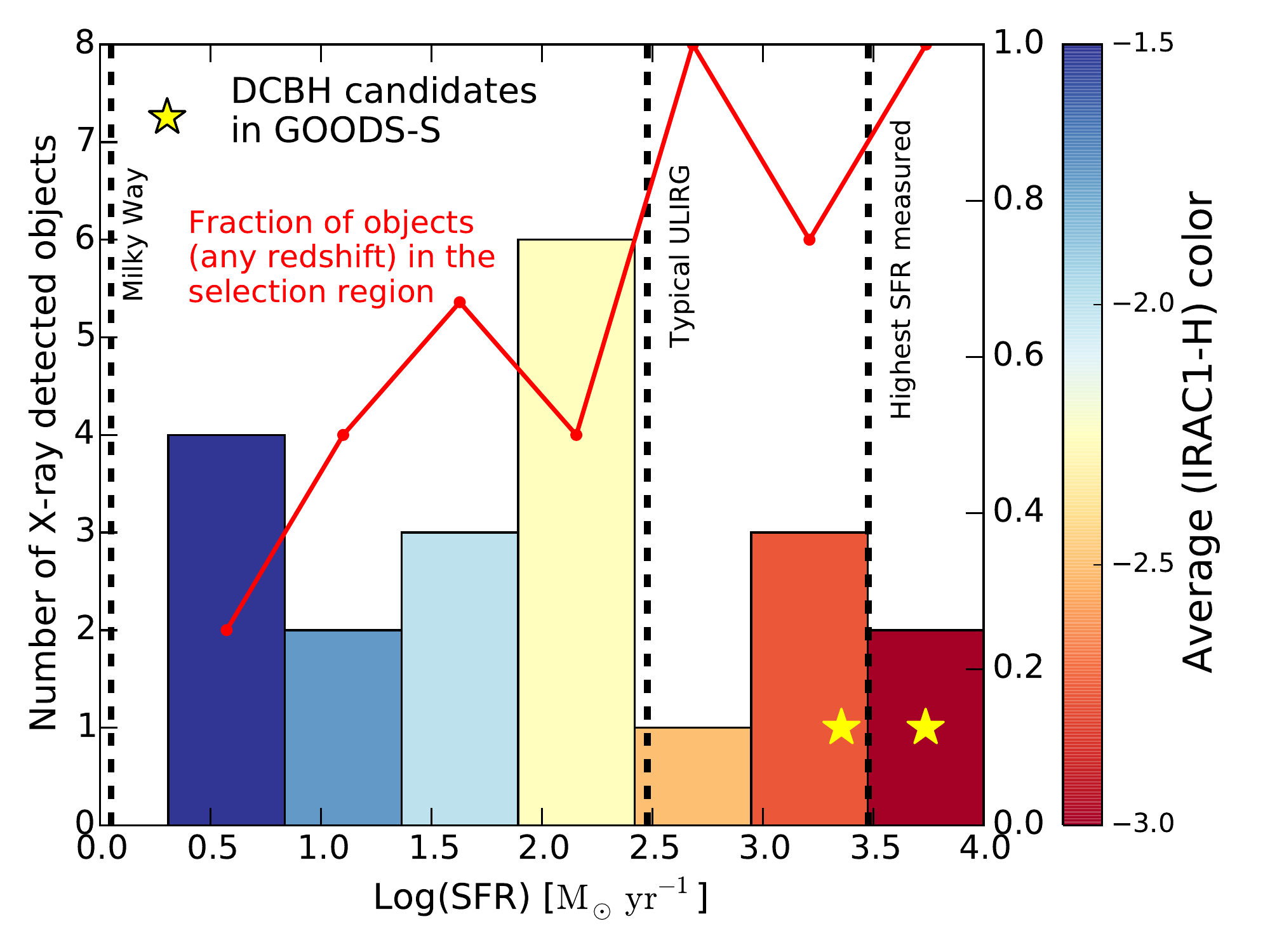}
\caption{Distribution of star formation rates of the X-ray detected objects in GOODS-S (excluding object $28476$, whose X-ray detection is still debated). Each column of the histogram is colored according to the average (IRAC1-H) color index of the objects falling within the SFR bin, as shown in the color bar. SFRs for the Milky Way \citep{Robitaille_2010}, a typical ULIRG \citep{Lefloch_2005} and for a massive maximum-starburst galaxy at $z=6.34$ \citep{Riechers_2013} are shown for comparison. The fraction of objects in each bin, at any redshift, falling into our selection region is shown as a red line, on the right vertical axis. The two DCBH candidates selected by our method are shown with yellow stars.}
\label{fig:SFR_hist}
\end{center}
\end{figure}

Fig. \ref{fig:SFR_hist} shows the distribution of star formation rates (SFRs) of the X-ray detected objects in GOODS-S (excluding object $28476$, whose X-ray detection is still debated).
An increase in the SFR corresponds to a more negative value of the average (IRAC1-H) color index, as the color bar shows. As expected, a redder photometry is in general due to a larger dust extinction, which requires star formation to be active in the host galaxy. Nonetheless, fitting the photometry of extremely red spectra, as the ones of our two DCBH candidates (shown as yellow stars) requires SFRs comparable to or larger than the highest ever measured, for a massive maximum-starburst galaxy at $z=6.34$ (\citealt{Riechers_2013}, see also \citealt{Barger_2014} where the authors find a characteristic maximum SFR of $\sim 2000 \, \mathrm{\Msun \, yr^{-1}}$). The computed SFR for object $29323$ is $\sim 2$ times larger than the maximum SFR ever measured, $\sim 20$ times larger than the typical SFR of ULIRGs \citep{Lefloch_2005} and $\sim 5000$ times the SFR of our Galaxy \citep{Robitaille_2010}. Instead of evoking extremely large and unrealistic SFRs, we propose that these objects host a central DCBH (see Fig. \ref{fig:seds} where the photometry of object 29323 is very well fitted by our DCBH model). The redness of their spectra is explained by the large, absorbing gas column densities in the host halo implicit in the DCBH evolution scenario.

A final remark: at low redshifts and for mildly negative color indexes, the colors of DCBHs may be mimicked by star forming galaxies. Fig. \ref{fig:SFR_hist} shows with a red line the fraction of X-ray detected objects at any redshift that fall in our selection region. It is likely that the redness of some of these spectra may be genuinely explained by a burst of star formation underway in the host galaxy. At high redshifts and for very negative color indexes (IRAC1-H $\lesssim -2.5$), the fraction of objects selected by our method is close or equal to unity and the required SFRs become unrealistic. We suggest that these objects are most likely high-$z$ black hole seeds.

%%%%%%%%%%%%%%%%%%%%%%%%%%%%%%%%%%%%%%%%%%%%%%%%%%%%%%%%%%%%%%%%%%%%%%
%% SECTION 5: Caveats
%%%%%%%%%%%%%%%%%%%%%%%%%%%%%%%%%%%%%%%%%%%%%%%%%%%%%%%%%%%%%%%%%%%%%%
\section{Caveats}
\label{sec:caveats}

In this Section we revise some of the most important caveats of the current work, related to the assumptions of our radiation-hydrodynamic code and to the uncertainties in the photometric redshift estimate of the X-ray detected sources.

\subsection{Assumptions of the radiation-hydrodynamic code}
As described in Section \ref{sec:implementation}, our radiation-hydrodynamic code simulates the accretion onto an already formed DCBH with some assumptions on the geometry of the flow and on the environment hosting the compact object.

On sufficiently large scales (comparable with the Bondi radius of the DCBH) the geometry is assumed to be spherical, with gas moving radially without angular momentum. As discussed in \cite{PVF_2015}, this assumption is supported by several studies (e.g. \citealt{Choi_2013,Choi_2015}) on the triaxiality of primordial halos, which should allow the gas to flow radially well beyond the centrifugal radius of the system. 

The accretion rate is computed self-consistently from the density and velocity fields of the gas at the inner boundary of our simulation domain. While some accretion models allow for a fraction of the inflowing mass to be lost on small scales during the accretion process (see e.g. \citealt{Begelman_2012, Coughlin_2014}), our simulations assume that the gas reaching the inner boundary of our spatial domain is entirely accreted onto the black hole. The bolometric luminosity emerging from the accretion disk is then computed consistently from the accretion rate (see Eq. \ref{l_bol}).

Star formation is not active in our simulations.
Consequently, the emerging spectrum is computed for a metal-free environment. 
The presence of a stellar component in the host halo (Pop II or Pop III) may modify the emerging spectrum in the optical/UV region, while we checked that the infrared part, upon which our photometric method is based, is totally unaffected. Note, however, that, in the classical DCBH scenario, fragmentation, and hence star formation, is heavily suppressed by the external UV background.

\subsection{Redshift estimates for DCBH candidates}
\label{subsec:caveats_red}
The redshift estimates for sources 33160 and 29323, selected by our method as DCBH candidates (see Fig. \ref{fig:colors_highz}), are photometric and subject to uncertainties (see Fig. \ref{fig:pdf}, right panel).

As already mentioned, the photometric redshifts of the GOODS-S galaxies have been computed with the unconventional technique described in \cite{Dahlen_2013}. In summary, it is an optimal bayesian combination of six different probability distribution functions of the photometric redshift computed by six different groups within the CANDELS Team. For the object 29323, the individual probability distribution functions are centered at $z \sim 6.5$ and at $z \sim 9.7$, and only one solution out of six gives $z \approx 10$. The $68\%$ ($95\%$) confidence level region is between $z_{min}=5.46$ and $z_{max}=9.86$ ($z_{min}=2.36$, $z_{max}=9.97$). From the final probability distribution function for this object we can conclude that the redshift is unconstrained at $z \gtrsim 5.5$ at $1\sigma$ (or $2.4$ at $2\sigma$), due to the power law SED without significant breaks at the bluer wavelengths.
Given the extreme steepness of the infrared SED of object 29323 (see Fig. \ref{fig:seds}), however, it would still be selected as a DCBH candidate down to $z \sim 2$.

Other photometric redshift catalogs on the same field have been published (e.g. by the 3D-HST collaboration, \citealt{Brammer_2012}, and by \citealt{Hsu_2014}). Nonetheless, we decided to adopt the CANDELS photometric redshifts for several reasons. 
Firstly, the 3D-HST photometric redshifts are very accurate for bright objects with strong emission lines. For faint galaxies such as the ones in this paper we prefer to use the CANDELS photometric redshifts, which are based on point spread function deconvolved photometry, which is quite robust also for faint and contaminated objects (e.g. object 29323). Moreover, the CANDELS photometric redshifts have the unique feature of averaging the differences between different photometric redshift codes, which could have their own biases.
To conclude, the accuracy in the determination of the photometric redshift at $z \gtrsim 4$ relies mainly on the absorption of the intergalactic medium, which is an accurate feature independently of the adopted templates.
The photometric redshifts of our DCBH candidates are $z(29323) \approx 5.3$ and $z(33160) \approx 4.9$ in the 3D-HST catalog and $z(29323) \approx 4.7$ and $z(33160) \approx 3.4$ in the \cite{Hsu_2014} catalog.

%%%%%%%%%%%%%%%%%%%%%%%%%%%%%%%%%%%%%%%%%%%%%%%%%%%%%%%%%%%%%%%%%%%%%%
%% SECTION 6: Discussion and Conclusions
%%%%%%%%%%%%%%%%%%%%%%%%%%%%%%%%%%%%%%%%%%%%%%%%%%%%%%%%%%%%%%%%%%%%%%
\section{Discussion and Conclusions}
\label{sec:disc_concl}
Supported by radiation-hydrodynamic simulations, we have devised a novel method to identify black hole seed candidates at $z \gtrsim 6$: at high-$z$ the infrared SED of pristine DCBHs is predicted to be significantly red. At lower redshifts, intervening processes (e.g. star formation, metal pollution, outflows) make the selection more uncertain. By applying our method to state-of-the-art photometric observations of the CANDELS/GOODS-S field, we select the only $2$ sources at $z \gtrsim 6$ with a robust X-ray detection. The SFRs required to mimic their extremely red spectra are unrealistic ($\gtrsim 2000 \, \mathrm{\Msun \, yr^{-1}}$), being comparable to or larger than the highest SFR ever measured, in a maximum-starburst galaxy. \textit{These objects represent the most stringent observational identification of black hole seed candidates, likely formed as DCBHs, so far obtained}. The possible presence of markers of DCBH activity, such as the predicted He II emission line, needs to be investigated with spectroscopic follow-up observations.

The extension of the current work to more infrared photometric bands may prove to be useful in distinguishing the different classes of objects observed, especially at lower redshifts. However, our two-dimensional color-color plot aims at conveying the simple and clear idea that, in looking for pristine DCBH candidates, observational efforts should be focused on objects with very steep infrared spectra.

This work extends previous efforts in understanding the observational features of DCBHs.
Our previous studies \citep{PFVD_2015,Pallottini_Pacucci_2015_CR7} investigated the spectrum emitted from these sources, foreseeing the possibility that they could be observed by the JWST. We suggested that CR7, the brightest Ly-$\alpha$ emitter discovered to date, could be powered by a typical DCBH of initial mass $\sim 10^5 \, \mathrm{\Msun}$. Recently, CR7 has been the subject of great interest in the community, with a large wealth of studies investigating its nature (e.g. \citealt{Agarwal_2015, Hartwig_2015, Visbal_2016, Smith_2016, Smidt_2016, Dijkstra_2016_CR7}), most of them confirming the possibility that this source may host a DCBH.
The efforts to observe these objects have recently spread to other regions of the electromagnetic spectrum. For instance, \cite{Dijkstra_2016} suggested that the large neutral hydrogen column densities of primordial gas at $\sim 10^4 \, \mathrm{K}$ with low molecular abundance (all key aspects of the DCBH scenario) give rise to stimulated fine structure emission at $\sim 3 \, \mathrm{cm}$ in the rest-frame. Detecting this signal is challenging, but would provide direct evidence for the DCBH scenario.

Our current work has, of course, some limitations. Firstly, high-$z$ sources with a robust X-ray detection in the GOODS-S field are few: making a precise assessment on the robustness of our selection criterion is beyond current reach. Secondly, only deep high/medium-resolution spectroscopy may conclusively confirm the DCBH nature of these objects. A spectroscopic analysis of these sources is beyond the capabilities of current observatories (e.g. HST, VLT, Keck) because they are too faint. 
With the start of JWST operations, a large wealth of infrared spectroscopic data (extended above the $\sim 2.5 \, \mathrm{\mu m}$ limit of ground telescopes) will be available, disclosing to our eyes the first glimpses of light in the Universe. 
Currently, we probe only the most massive and luminous black holes, the peak of their mass distribution. JWST will mark a breakthrough in this field, by detecting the light from the most distant stars and accreting black holes, probing the mass range ($10^{4-5}\, \mathrm{\Msun}$) of the first black hole seeds, if they were formed via the DCBH scenario.
Our work establishes a solid theoretical framework to interpret these data with the aim of finding the first black holes in the Universe.

\vspace{0.2cm}
We thank N. Cappelluti, C. Feruglio, S. Gallerani, A. Loeb and A. Mesinger for helpful discussions and comments.
This work is based on observations taken by the CANDELS Multi-Cycle Treasury Program with the NASA/ESA HST.
Observations were also carried out using the VLT, the ESO Paranal Observatory and the Spitzer Telescope.

%%%%%%%%%%%%%%%%%%%%%%%%%%%%%%%%%%%%%%%%%%%%%%%%%%%%%%%%%%%%%%%%%%%%%%
%%  REFERENCES
%%%%%%%%%%%%%%%%%%%%%%%%%%%%%%%%%%%%%%%%%%%%%%%%%%%%%%%%%%%%%%%%%%%%%% 

\bibliographystyle{mnras}
\bibliography{ms}

\label{lastpage}
\end{document}